\documentclass[galaxies,review,accept,moreauthors,pdftex,10pt,a4paper]{mdpi}
\usepackage{amssymb}
\usepackage{aas_macros}
\usepackage[labelformat=simple]{subfig}

\firstpage{1} 
\pubvolume{6}
\issuenum{4}
\articlenumber{135}
\pubyear{2018}
\copyrightyear{2018}
\history{Received: 7 November 2018; Accepted: 4 December 2018; Published: 6 December 2018}
\updates{yes} 

\theoremstyle{mdpi}
\newcounter{thm}
\newcounter{ex}
\newcounter{re}
\Title{Flux Distribution of Gamma-Ray Emission in Blazars: The Example of Mrk~501}

\Author{Carlo Romoli ${^{1,}}$*, Nachiketa Chakraborty ${^{1,}}$*, Daniela Dorner ${^{2,}}$*, Andrew~M.~Taylor ${^{3,}}$* and Michael~Blank~${^{2,}}$* for the FACT and H.E.S.S. Collaborations}
\address{%
$^{1}$ \quad Max-Planck-Institut f\"ur Kernphysik, P.O. Box 103980, D 69029 Heidelberg, Germany \\
$^{2}$ \quad Universit\"at W\"urzburg, Institute for Theoretical Physics and Astrophysics Emil-Fischer-Str. 31,
97074~Würzburg, Germany\\
$^{3}$ \quad Deutsches Elektronen-Synchrotron (DESY), D-15738 Zeuthen, Germany}

\corres{Correspondence: cromoli@mpi-hd.mpg.de (C.R.); cnachi@mpi-hd.mpg.de (N.C.); dorner@astro.uni-wuerzburg.de (D.D.); andrew.taylor@desy.de (A.M.T.); michael.blank@stud-mail.uni-wuerzburg.de (M.B.)}

\abstract{Flux distribution is an important tool to understand the variability processes in active galactic nuclei. We now have available a great deal of observational evidences pointing towards the presence of log-normal components in the high energy light curves, and different models have been proposed to explain these data.
Here, we collect some of the recent developments on this topic using the well-known blazar Mrk~501 as example of complex and interesting aspects coming from its flux distribution in different energy ranges and at different timescales. The observational data we refer to are those collected in a complementary manner by Fermi-LAT over multiple years, and by the First G-APD Cherenkov Telescope (FACT) telescope and the H.E.S.S. array in correspondence of the bright flare of June 2014.}


\begin{document}



\section{Introduction}


The origin of variability in the emission across the electromagnetic spectrum in active galactic nuclei (AGNs), especially at gamma-ray energies, is still a puzzle due to the lack of clear understanding of the emission mechanisms leading to the high energy flux.

A crucial improvement is the increase in sensitivity at TeV energy that happened in the past 20~years with the present generation of Imaging Air Cherenkov Telescopes (IACTs), like e.g.,\ H.E.S.S., which allowed high statistics measurements of bright flares. Another important development is the observing strategy of high cadence monitoring of the brightest AGNs as now carried out with instruments like First G-APD Cherenkov Telescope (FACT) and HAWC. All these aspects are making it possible for TeV astronomy to move towards a more mature phase.
Especially in the context of multi-wavelength observations, the~long-term monitoring programs to have continuous observations of variable sources are very useful.
This unprecedented coverage on wide spectral and temporal ranges is shedding new light on the phenomenon of variability.

One of the key properties is the flux distribution (often indicated as PDF: Probability density function) of the sources.
A Gaussian (or ``normal'') flux distribution would be an indication of a linear summation of components contributing to building up the emission. Log-normal distributions (distributions which are Gaussian in the logarithm of the flux) are instead naturally obtained via multiplicative (or cascade)---like processes see e.g., \cite[][]{2005MNRAS.359..345U} and references therein.
Therefore, the PDF highlights the mathematical form and the class of mechanisms producing variability in the observed emission. In particular, it can rule out entire classes of additive multi-zone models.

The next subsections will go in more details into to the topic of flux distributions in AGNs at high and very high energy and will set up the context of the analysis on the AGN Mrk~501 presented here.

\subsection{Flux Distributions of Blazars}
\label{sec:fldist}

Most of the gamma-ray light curves available for AGNs show a departure from a Gaussian distribution with a tail of high fluxes \cite{2015ApJ...810...14A}. Historically, this behaviour was found in the X-ray light curves for different classes of AGNs see e.g., \cite[][]{2005MNRAS.359..345U,2009A&A...503..797G}.
The log-normal behaviour at X-ray energies observed here is usually explained in terms of fluctuations in the Shakura-Sunyaev \cite{1976MNRAS.175..613S} accretion disk  model. While this model is used to explain log-normality at other wavelengths assuming these fluctuations are imprinted on emission from the jet as well, fast variations at gamma-ray energies with a log-normal PDF cannot be explained. This is because fluctuations from the accretion disk cannot include the relativistic Doppler boosting needed to justify minute timescale variability. Furthermore, effects due to the change of the jet viewing angle can affect the variability distribution, due to changes in the Doppler boosting.
Thus, gamma rays are crucial for studying log-normality.

\subsubsection{Observational Evidences in Gamma Rays}
The advent of high-sensitivity, ground-based gamma-ray telescopes (like MAGIC, H.E.S.S., and~VERITAS) enables the characterization of the flux distribution also in the very high energy (VHE; $E>100$ GeV) regime.
One of the most famous examples in this regard is the blazar PKS~2155-304. This~source is regularly observed by the H.E.S.S. experiment and underwent an extremely bright flaring activity in 2006 \cite{PhysRevLett.101.170402}.
A detailed study of long-term data collected by the H.E.S.S. telescopes has shown clearly the presence of log-normal behaviour in the light curve, both in the ``quiescent'' state and in the ``flaring'' state, as shown in Figure~3 of \cite{2010A&A...520A..83H}.
A~further analysis involving a much larger low state dataset solidly confirmed this trend \cite{2017A&A...598A..39H} which is seen also at different wavelengths \cite{2015arXiv150903104C}.
A~study including 18~years of public data of Mrk\,421 also revealed a log-normal behaviour fitting the flux distribution with the sum of a Gaussian and a log-normal distribution  \cite{2010A&A...524A..48T}.
In 2011, the observation of the gamma-ray sky has seen the arrival of another player: The FACT\footnote{\url{https://www.fact-project.org}}. Despite~being a small telescope (mirror area of $\sim$9.5\,m$^2$), it is equipped with a silicon photo-detectors instead of the traditional photo multipliers tubes. This camera choice allows it to operate in much higher external luminosity conditions (e.g.,~moon time) increasing its monitoring capabilities. Furthermore, its observation strategy is focused on the monitoring of a small sample of sources, mostly AGNs, to build unbiased light curves and explore in a more effective way the variability of TeV sources. Using these unbiased observations, the departure from ``normality'' in the PDF is confirmed for other sources \cite{2017ICRC...35..609D}.

At GeV energies, the field is explored using the {\it Fermi}-LAT telescope, which has been scanning the high energy (HE; $E>100$ MeV) sky since August 2008, in principle able to observe gamma rays with an energy going from few tens of MeV to hundreds of GeV (where the performance becomes limited by the low photon flux). Thanks to its constant monitoring, a great deal of data is available to study the statistical properties of AGN light curves.
Studies on the brightest blazars seen by the LAT show the same trend when the light curves are binned on a monthly timescale \cite{2018arXiv180504675S}. However, the {\it Fermi}-LAT, having a collection area of $\sim$1 m$^2$, lacks the necessary sensitivity to perform a detailed study also on fainter objects, where it can become hard to gain enough statistics on short timescales to perform these kind of studies, as shown in Section~\ref{sec:results}.

\subsubsection{Models to Explain Log-Normality}
The log-normal behaviour can be naturally explained by the effect of a multiplicative process and several models attempt to explain this phenomenon. Uttley et al. \cite{2005MNRAS.359..345U} have linked the presence of log-normality to non-linear variability aspects which could be explained by variations in the accretion flow that get coupled together and that are then transmitted to the emission region. In~the Shakura-Sunyaev type accretion disk, fluctuations in the viscosity parameter propagating from the outer rings to the inner rings in the accretion disk can naturally produce a log-normal PDF. These~models have also the advantage to be able to produce a variability power spectrum characterized by a power law \cite{1997MNRAS.292..679L}; a property often seen for AGNs, not only at gamma-ray energies, but at other wavelengths see e.g., \cite[][]{1987Natur.325..694L, 2017A&A...598A..39H, 2018ApJ...863..175G}.
The presence of log-normality however does not completely rule out additive models, but makes them less natural: \citet{2012A&A...548A.123B} have shown that it is possible to obtain a flux distribution that would resemble a log-normal one by the addition of the emission of a large ($O(10^4)$) number of  ``mini-jets'' randomly oriented inside the main AGN outflow. The~advantage of this model would be to have a direct connection between the source of variability and the emission from the jet.
Skewness in the PDF can also be obtained in the case of perturbations in the acceleration and/or cooling timescales in the emission region \cite{2018MNRAS.480L.116S}. In this scenario, the skewness of the flux distribution is due to Gaussian perturbations in the acceleration and/or escape timescales of the primary particles in the acceleration region.
While fluctuations in the acceleration rate would yield true log-normal distributions, fluctuations in the escape rate will be associated to a distribution that is neither normal nor log-normal, but that still presents skewness toward high fluxes and a mild linear correlation between flux and excess variance.
An important prediction of this model is that the skewness of the flux distribution is more prominent at high energies, while tend to disappear at low energy \cite{2018MNRAS.480L.116S}. While promising, this model refers to the steady state of the emission and furthermore it has yet to be properly tested using more realistic parameters encountered in the modelling of AGNs.

\subsubsection{Importance of Simulations}
A key aspect of testing models producing log-normal PDFs is to firmly establish statistically that the observed distribution is indeed log-normal and not simply an artefact of observations. This is achieved by simulating light curves with a known PDF.
The correct procedure requires to produce sets of simulated light curves taking into account all the observational constraints as the sampling frequency and gaps in the data sets.
We use the approach from \cite{CB2018} which modifies the Timmer-Koenig method of simulating light curves \cite{1995A&A...300..707T} with a power-law power spectral density PSD and a Gaussian PDF, but additionally we fold in cadence of the actual observations. This is implemented in as follows. We first simulate at least 1000 Timmer-Koenig Gaussian light curves with with PSD indices 1 and 2 ; as we find, the results are not sensitive to these values which are representative of the range of observed estimates for blazars. However, in the case with PSD indices $>$1, the total power diverges at low frequencies and hence there is a violation of strict stationarity with consequent departures from Gaussianity, the likes of which are demonstrated in \cite{2003MNRAS.345.1271V}. In reality, the PSD at lower frequencies and longer timescales is less steeper ($<$1) as the total power is finite. Detailed exploration of effects of this on the estimated PDF is beyond the scope of this proceedings and will appear in \cite{Morrisetal}.

These light curves have bin-size at least as small as the observed bin-size, if not smaller by a small factor. The length of the time series is 5 times longer which reduces the problem of red noise leakage. These are "idealised" light curves that have a mean of 0 and variance of 1. We then impose on these light curves the observational properties. This includes shifting and scaling the flux values to match the observed mean and variance. We introduce gaps in the simulated light curve at the exact times where there are gaps in the observed light curves. This folds in the uncertainty due to irregularity in cadence. The resultant light curves are what maybe considered ``realistic, artificial light curves''. This~artificial ensemble allows us to statistically test the occurrence of a tail or skew or a deviation from Gaussianity. We can take this further by simulating log-normal light curves to test consistency with a true log-normal distribution.

\subsection{Markarian 501}

The High-synchrotron peaked BL Lac (HBL) type object Markarian~501 (Mrk~501) is one of the most relevant AGN in the field of VHE astrophysics. It was discovered for the first time by the Whipple Observatory in 1996 \cite{1996ApJ...456L..83Q} (second extragalactic source detected at VHE), and it is one of the closest blazars, being at a redshift of $z=0.034$. Due to its proximity and brightness in this energy range, it~has been often target of observational campaigns since its discovery, involving several energy ranges see~e.g.,~\cite[][]{2011ApJ...727..129A}.

At GeV energies, it is clearly detected by the {\it Fermi}-LAT as a point-like source with a hard energy spectrum (photon index of the power law model, $\Gamma < 2$) at energies above 100 MeV \cite{2015ApJS..218...23A}. The source is characterized by a very high degree of variability, especially in correspondence of the peaks of its broadband Spectral Energy Distribution (SED): In the X-ray and gamma-ray energy range \cite{2011ApJ...727..129A,2018arXiv180804300A}. In~particular at VHE, the emission often presents strong flares reaching fluxes as high as 10 times the flux of the Crab nebula (Crab Unit; C.U.) at 1 TeV. The spectra reconstructed during these events are also particularly interesting, extending up to $\sim$20 TeV energy following a power-law function without any apparent sign of cut-off once the absorption from the Extragalactic Background Light (EBL) is taken into account \cite{1999A&A...342...69A,1999A&A...350...17D,2018arXiv180804300A,2017AIPC.1792e0019C}. For a more general overview on Mrk~501, see also \cite{PanequeMRK501}. Being a very active source, it offers a perfect test bench for the study of flux distributions.

\section{Variability and Flux Distributions for Mrk~501 in 2014}
\label{sec:results}

\begin{figure}[H]
\centering
\subfloat[Fermi-LAT, monthly light curve]{\includegraphics[width=0.75\textwidth]{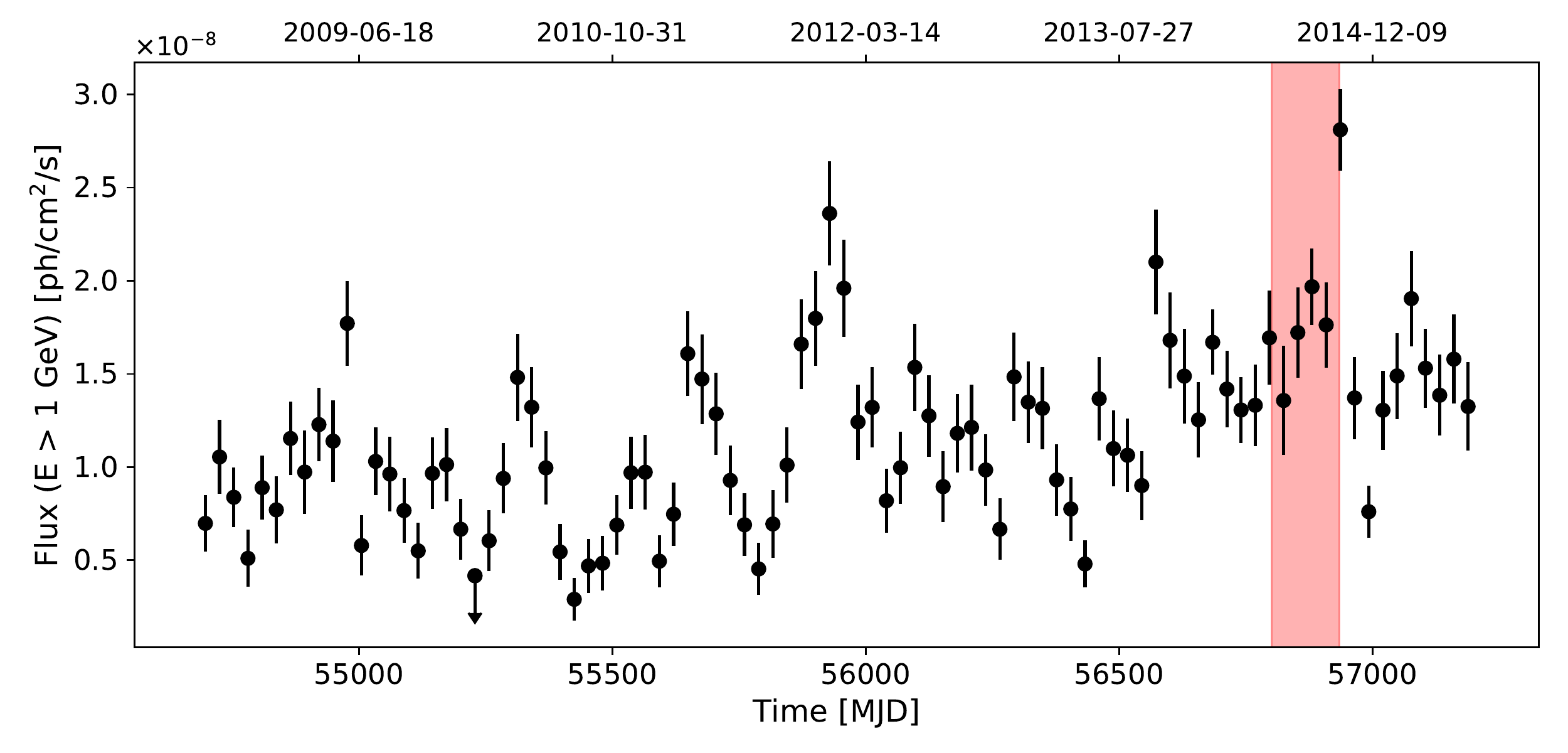}\label{fig:FERMImonth1gevLC}}\\
\subfloat[FACT, nightly light curve]{\includegraphics[width=0.75\textwidth]{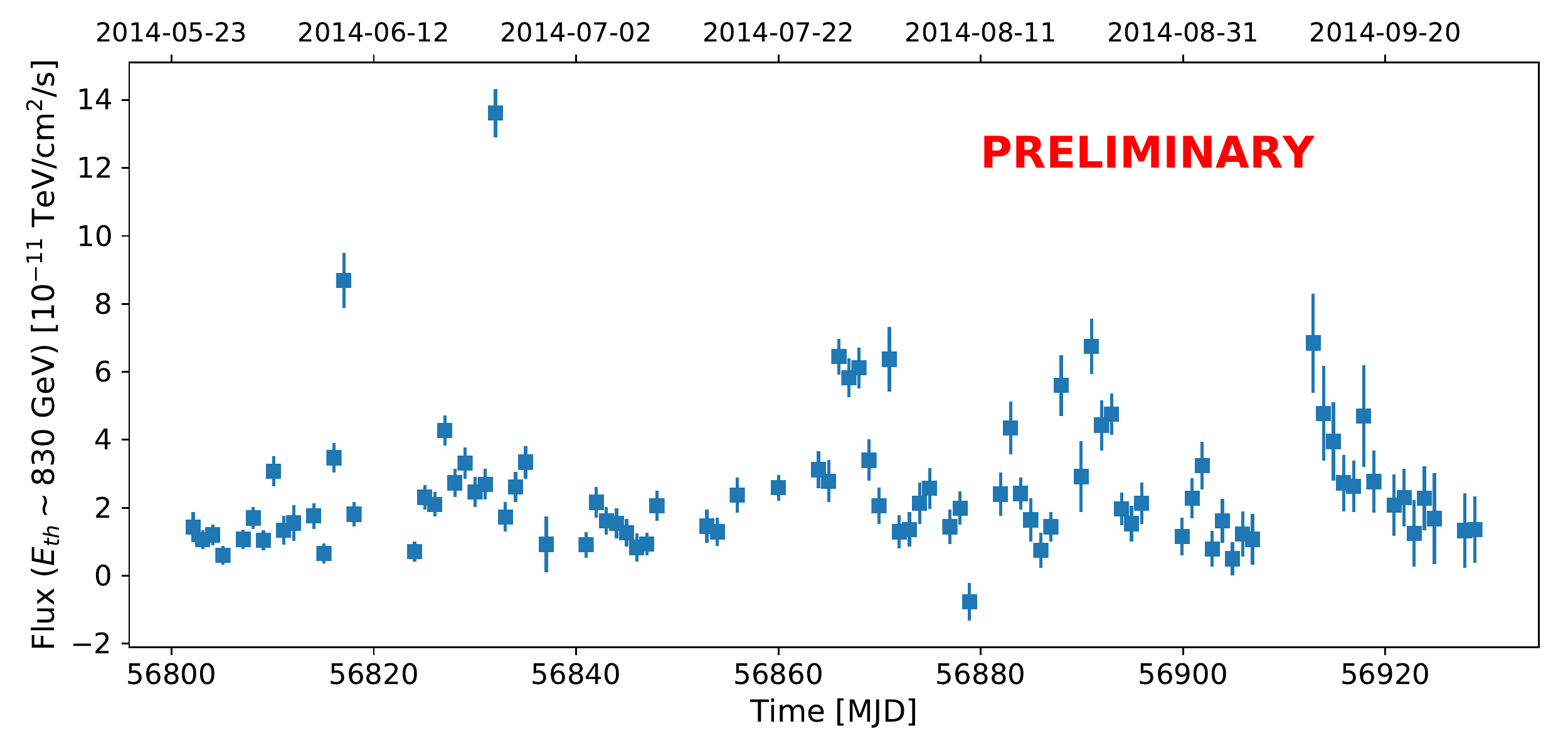}\label{fig:FACTnightLC}}
\caption{({\bf a}) Fermi-LAT photon flux light curve (in units of photons/cm$^2$/s) for photon energies $E>1$ GeV spanning over a period of seven years, using a bin size is 28 days. The red band indicates the time range of the FACT data. ({\bf b}) FACT energy flux light curve (in units of TeV/cm$^2$/s) with nightly binning with an energy threshold $E_{th}\sim830$ GeV. The light curve spans a time interval of almost five months.}
\label{fig:lightcurves}
\end{figure}

The dataset available for this work spans a wide range of timescales and energies, thanks to the complementarity of the instruments involved in the observation. The light curves extracted from this dataset are shown in Figure~\ref{fig:lightcurves}.
The event taken as a reference for this work is the high state detected by the FACT telescope in June 2014 which triggered subsequent H.E.S.S. observations \cite{2014ATel.6268....1S}.
These~observations resulted also in the detection of an extremely bright flare on the 23rd of June 2014, when  Mrk~501 reached a luminosity in the VHE band close to its historical maximum \cite{2017AIPC.1792e0019C}.

Due to the brightness of the event, it is possible to study the source behaviour on a wide range of timescales: The FACT telescope allows us to cover the emission in an unbiased way with a good nightly sampling with an energy threshold $E_{th} \sim 830$ GeV, shown in Figure~\ref{fig:lightcurves}b, over a period of almost five months. The light curve has been obtained based on the automatic quick-look analysis described in~\cite{FACTQLA}. In addition, a data quality selection based on the cosmic-ray rate \cite{2017ICRC...35..779H} has been applied resulting in a data sample of about 215 hours. The excess rates have been converted to fluxes using the Crab Nebula data as described in \cite{2017ICRC...35..612M}.
The flaring episode can in addition be studied in more details thanks to the H.E.S.S.\ observation which yielded a light curve with a 4-minute binning even though, due to the large zenith angle of the H.E.S.S.\ observation, the energy threshold is above 2 TeV. The light curves (also in comparison with FACT) can be found in Figure~1 of \cite[][]{2017ICRC...35..608D}.

At GeV energies, the data rely on the the {\it Fermi}-LAT and span more than seven years of observations, going from August 2008 to July 2015 (MET from 239557417 to 459661531). As Mrk~501 is not very bright in this energy range during this period, the data were binned on a 28-day interval, and they are selected to have an energy comprised between $E>1$ GeV and 500 GeV, to safely avoid contamination from neighbouring sources thanks to the better PSF of the instrument in this energy range. The light curve is presented in Figure~\ref{fig:lightcurves}a. It was obtained after a full binned likelihood fit of the data using the standard Science Tools version 10.0.5, PASS8 instrument response functions and SOURCE photon class in a square region of interest with 28 degrees of side, centred in the position of Mrk~501. The fit took into account all the sources within 25 degrees angular distance from the central source contained in the 3FGL catalogue \cite{2015ApJS..218...23A} plus the diffuse background models provided by the {\it Fermi}-LAT Collaboration\footnote{\url{https://fermi.gsfc.nasa.gov/ssc/data/access/lat/BackgroundModels.html}}.
The 28-day light curve was obtained by fixing all the spectral parameters of the model to those obtained in the full time interval fit, except for the normalization of the sources flagged as variable in the 3FGL catalogue and the spectral parameters of Mrk~501 to account for the variability of nearby sources.

\subsection{PDF at GeV Energies: {\it Fermi}-LAT}

In order to establish the statistical significance of a functional form of the flux PDF, we need to compare it with an ensemble of simulations. Artificially generated light curves with different stochasticities quantified in terms of variability power spectra using the Timmer and Koenig method for Gaussian PDFs and modifying it to generate log-normal PDFs \footnote{It can be seen that the index of the power spectrum density (PSD) does not influence much the flux distribution, at least for the indices tested here.}. The observed PDFs are compared with those for simulated light curves to produce confidence intervals on the flux bins for statistical testing. In the presented analysis, pink and red noise (PSD index 1 and 2 respectively) simulations are chosen for statistical testing.
The histogram binning was chosen such that its size would not be smaller than the average error on the flux values in each light curve bin. This allows to neglect possible features arising simply from statistical fluctuations of the light curve. To ease the comparison between real data and simulations in both normal and log-normal case, a feature rescaling was implemented so that all the flux distributions would have mean 0 and variance 1.

\begin{figure}[H]
\begin{center}
\includegraphics[width=0.45\textwidth]{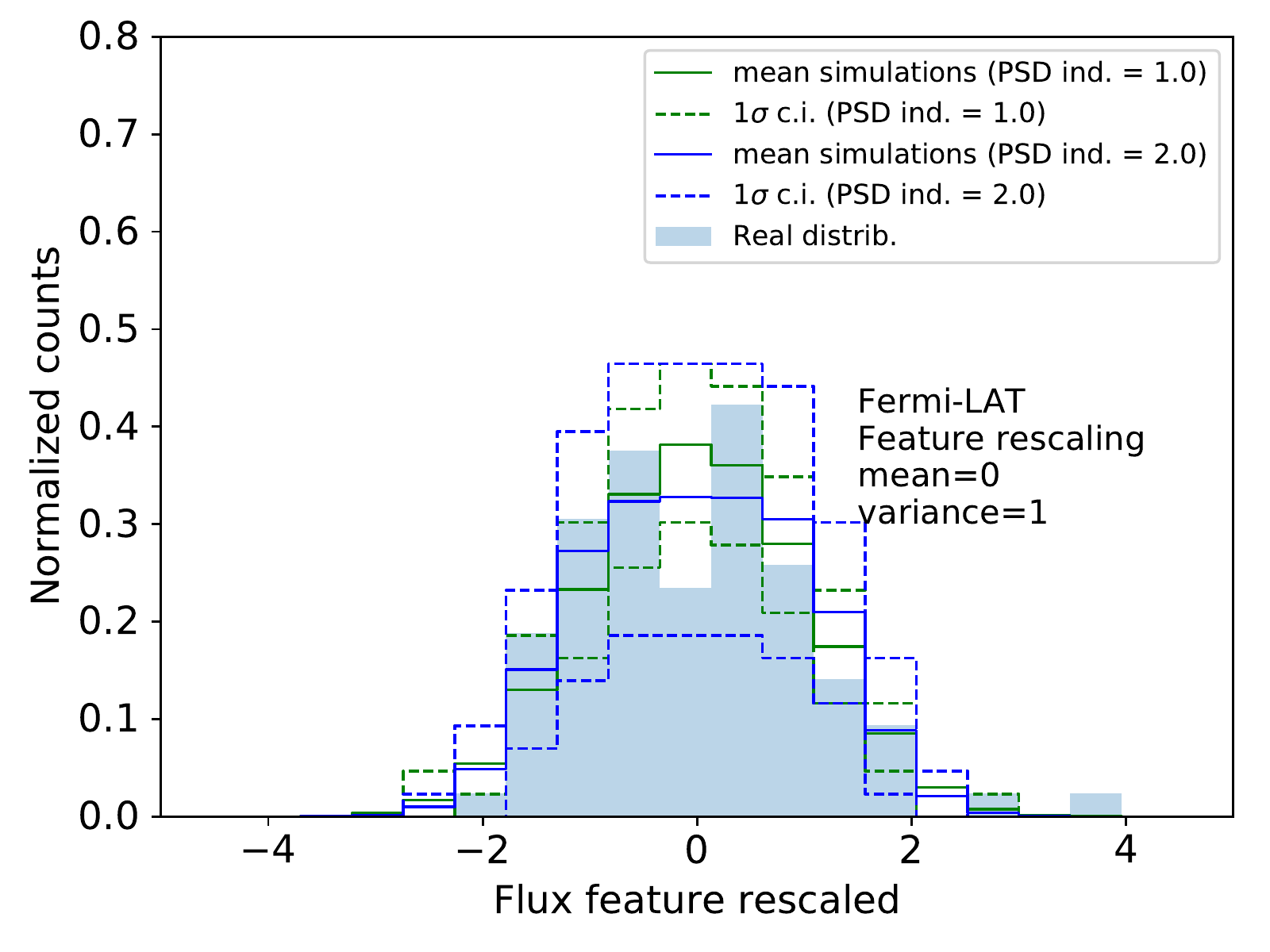}
\includegraphics[width=0.45\textwidth]{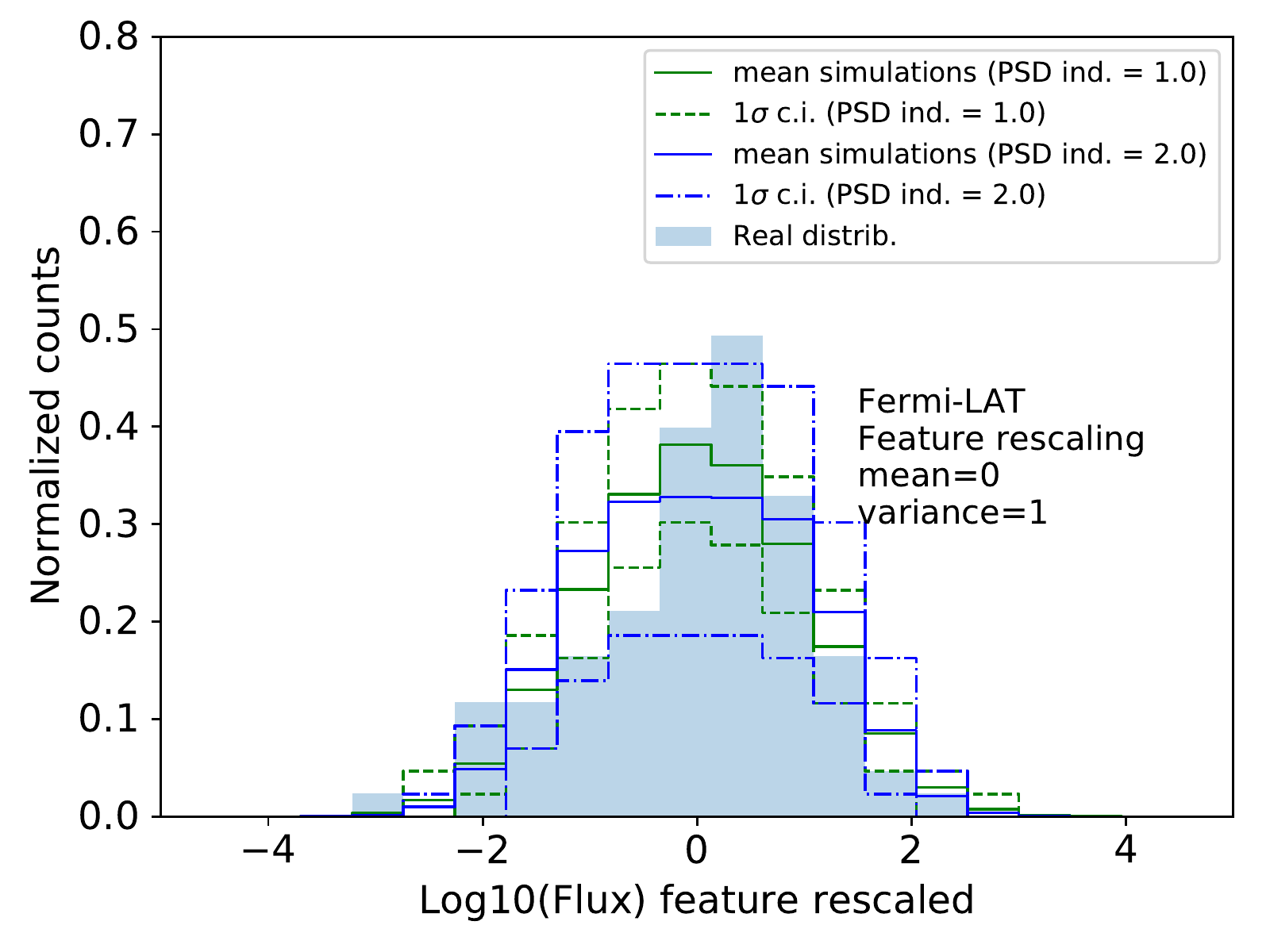}
\caption{The rescaled flux distribution of the long-term GeV {\it Fermi}-LAT light curves is shown here (for $E>1$ GeV) for the flux ({\bf right panel}) and the $\log_{10}$ of the flux ({\bf left panel}). The blue histogram is obtained from the real {\it Fermi}-LAT monthly light curve. The solid lines are the mean obtained from the light curve simulations with PSD index = 1 (green line) and PSD index = 2 (blue line). The dashed and dot-dashed lines represent instead the 1 sigma confidence intervals obtained by taking the 16{th} and 84{th} percentile of the distribution of the entries in the bins for the 5000 simulations adopted. These~results are robust to variations in binning and show no significant deviation between the {\it Fermi}-LAT data and a simulated normal distribution of the fluxes.}%
\label{fig:PDFFermi}
\end{center}
\end{figure}

As visible from Figure~\ref{fig:PDFFermi}, the distribution of the flux above 1 GeV, on timescales of months is  still consistent with a Gaussian distribution or a log-normal one. To statistically test the hypothesis, we~have run a Kolmogorov-Smirnov (KS) test to assess the similarities between the real data distribution and the simulated ones. Having a large number of simulated PDFs, we present the median and the 16th and 84th percentiles of the distribution of the \emph{p}-value of the test only for the case with PSD index = 1 (the results of this test are not strongly affected by the index value).

For the Gaussian simulations case, we obtain a \emph{p}-value of $0.80_{-0.15}^{+0.16}$ (where the super and sub scripts indicate the distance from the 84th and 16th percentile respectively). In the log-normal case, the~value is instead $0.68_{-0.34}^{+0.23}$. Because the KS test is biased to be more sensitive to differences in the centre of the distribution, we  ran the Shapiro-Wilks (SW) test to determine the degree of normality of the dataset independently from the simulations. This powerful test \cite{YapSim2011}
returned a 4\% probability for the GeV data to come from a pure normal distribution and a 13\% probability for the data to come from a pure log-normal distribution (testing the normality of the distribution of the $\log_{10}$ of the flux).

In \cite{2018arXiv180504675S}, the authors reject the normal distribution hypothesis through the Anderson-Darling test. However, their dataset spans a different energy range (using all the data above 100~MeV) and a different time range, which makes a direct comparison difficult. Our robust approach using both normality tests and comparison with simulations show the limited discriminatory power of the current estimates from GeV data in distinguishing between the normal and lognormal PDFs. In fact, this~highlights the importance of combining appropriate tests with simulations to make estimates.

The weekly light curves were also tested, to asses if a different behaviour arises when probing shorter timescales. However, because of the greater average uncertainty on the flux values for a shorter integration time, the histogram binning remains too coarse to draw any conclusion on a deviation from normality.

\subsection{PDF at TeV Energies, FACT and H.E.S.S.}

\begin{figure}[H]
\centering
\includegraphics[width=0.45\textwidth]{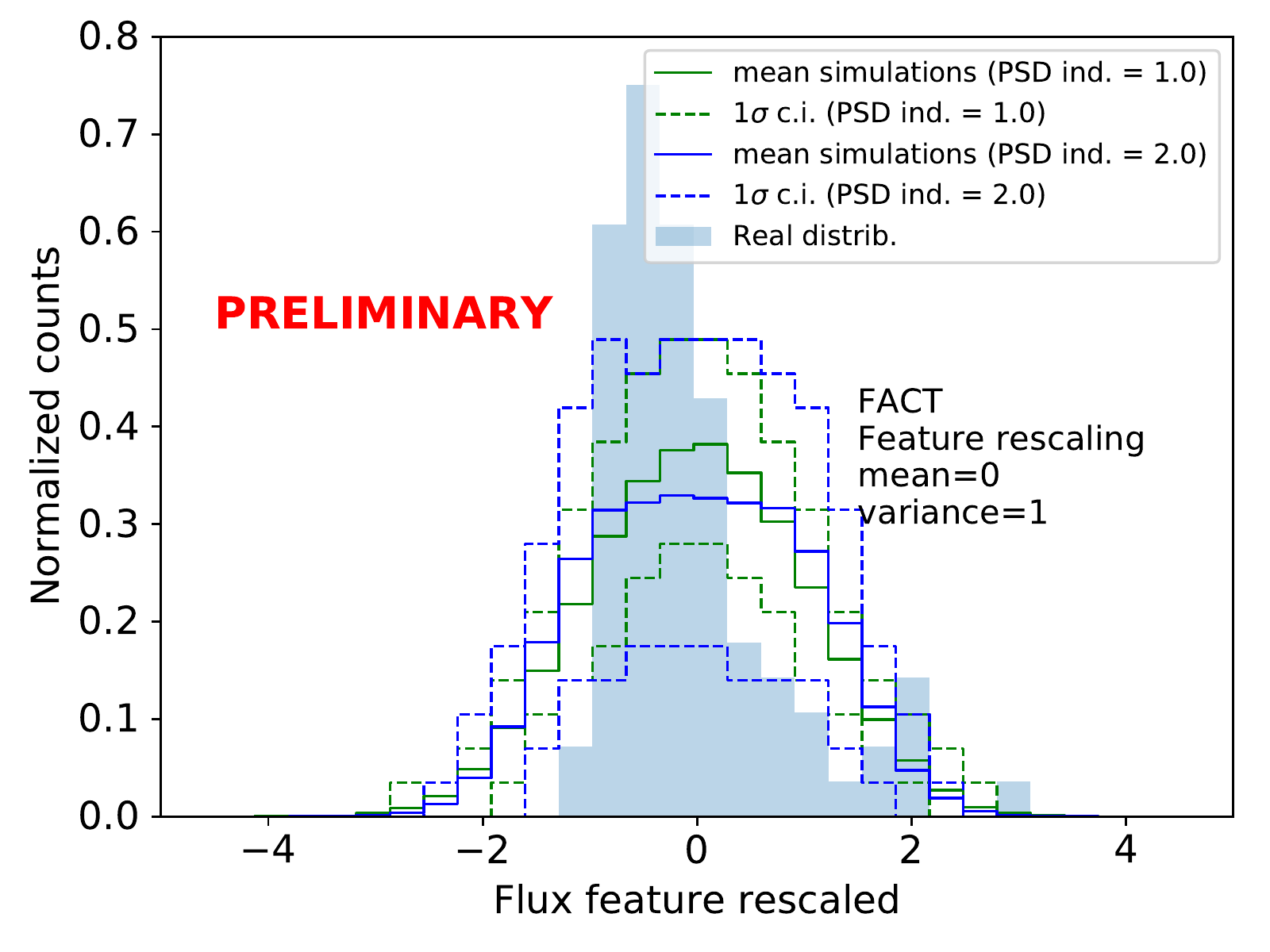}
\includegraphics[width=0.45\textwidth]{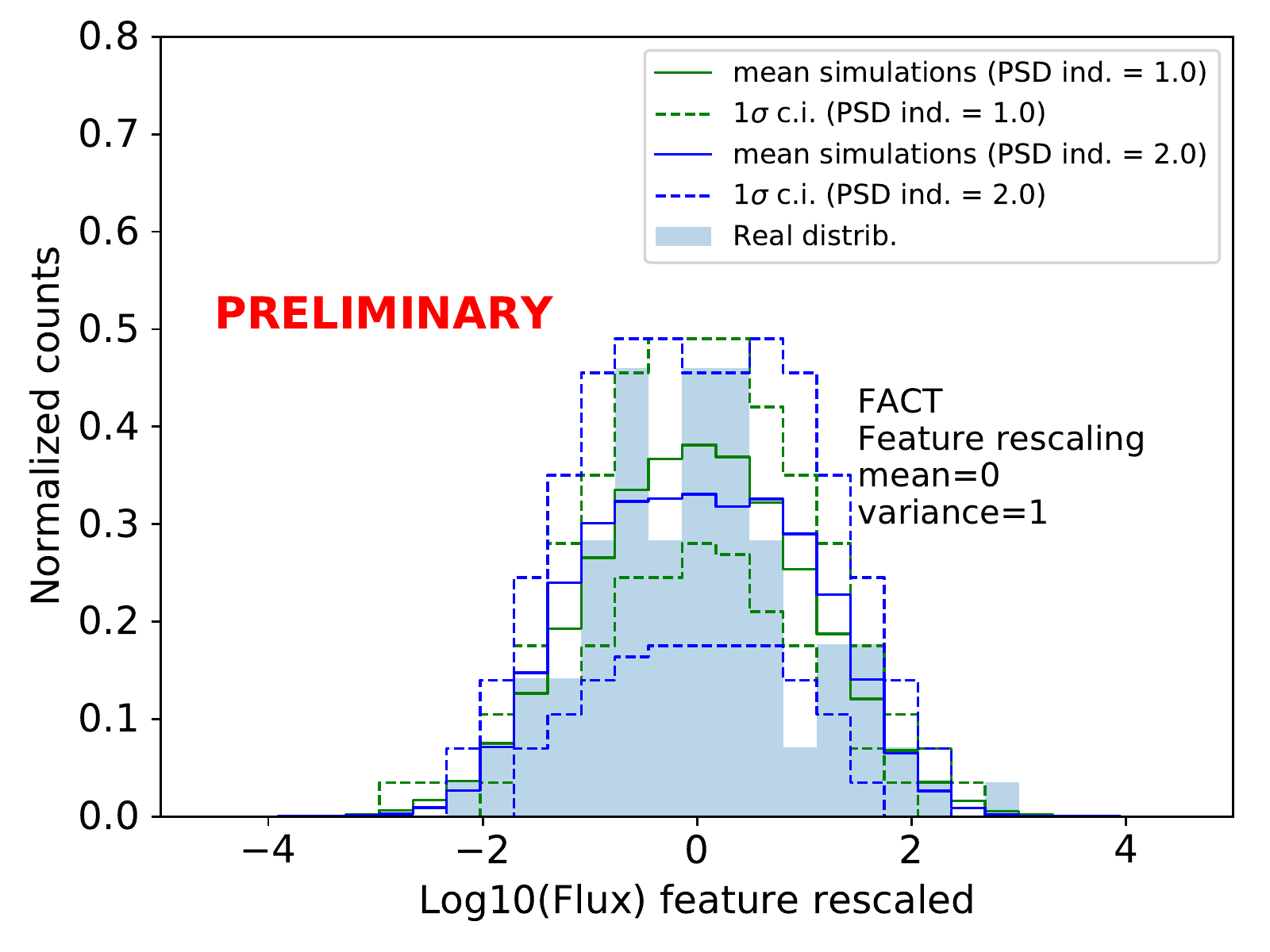}
\caption{Energy flux distribution of the FACT night-wise light curve compared with estimates from Gaussian ({\bf left panel}, 5000 simulations) and log-normal ({\bf right panel}, 1000 simulations) simulations. The colour codes are the same as in Figure~\ref{fig:PDFFermi}. In this case, the flux distribution is better reproduced using a log-normal simulated dataset.}
\label{fig:PDFFact}%
\end{figure}

In the VHE regime, the scenario is different and seems to be consistent across different timescales.
Figure~\ref{fig:PDFFact} illustrates the flux distribution obtained from the nightly light curve measured by FACT. Applying the same methodology used for the LAT light curve, it is possible to see an evident departure from a normal distribution, with the histogram skewed by the presence of tail towards higher fluxes.
Using simulated light curves rescaled to represent a log-normal behaviour, it is shown how the data distribution can be easily explained assuming log-normality in the source variability at these~timescales.

In this case, the data show a strong departure from a Gaussian distribution: The SW test yields a \emph{p}-value of $10^{-10}$ for the Gaussian hypothesis, while it is 0.73 for the log-normal one. The test using the simulated values also confirms the results, even though with less discriminative power: The normal hypothesis has a \emph{p}-value of $0.02^{-0.02}_{+0.10}$, while the log-normal reaches $0.43_{-0.31}^{+0.38}$.

\begin{figure}[H]
\centering
\includegraphics[width=0.45\textwidth]{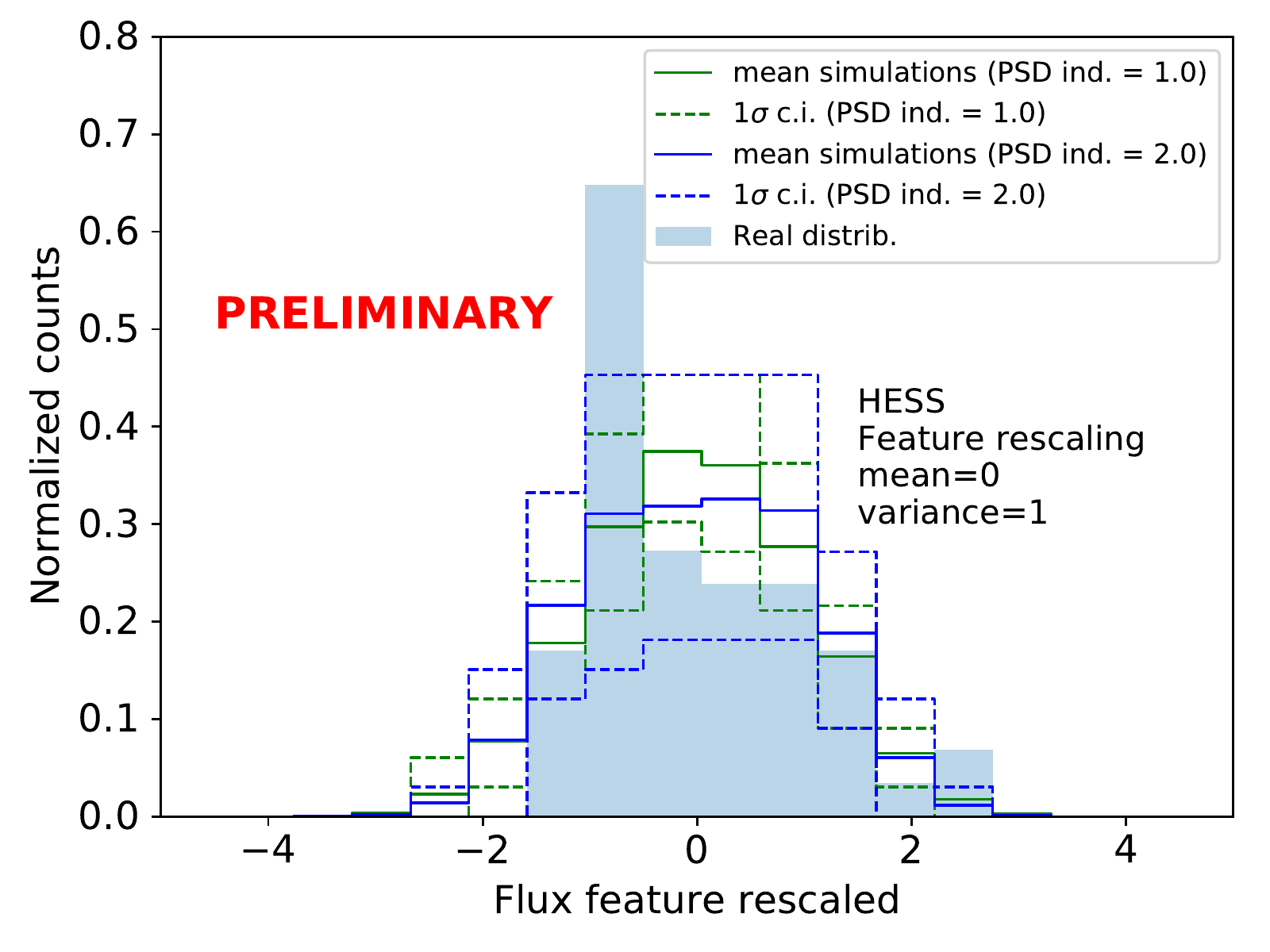}
\includegraphics[width=0.45\textwidth]{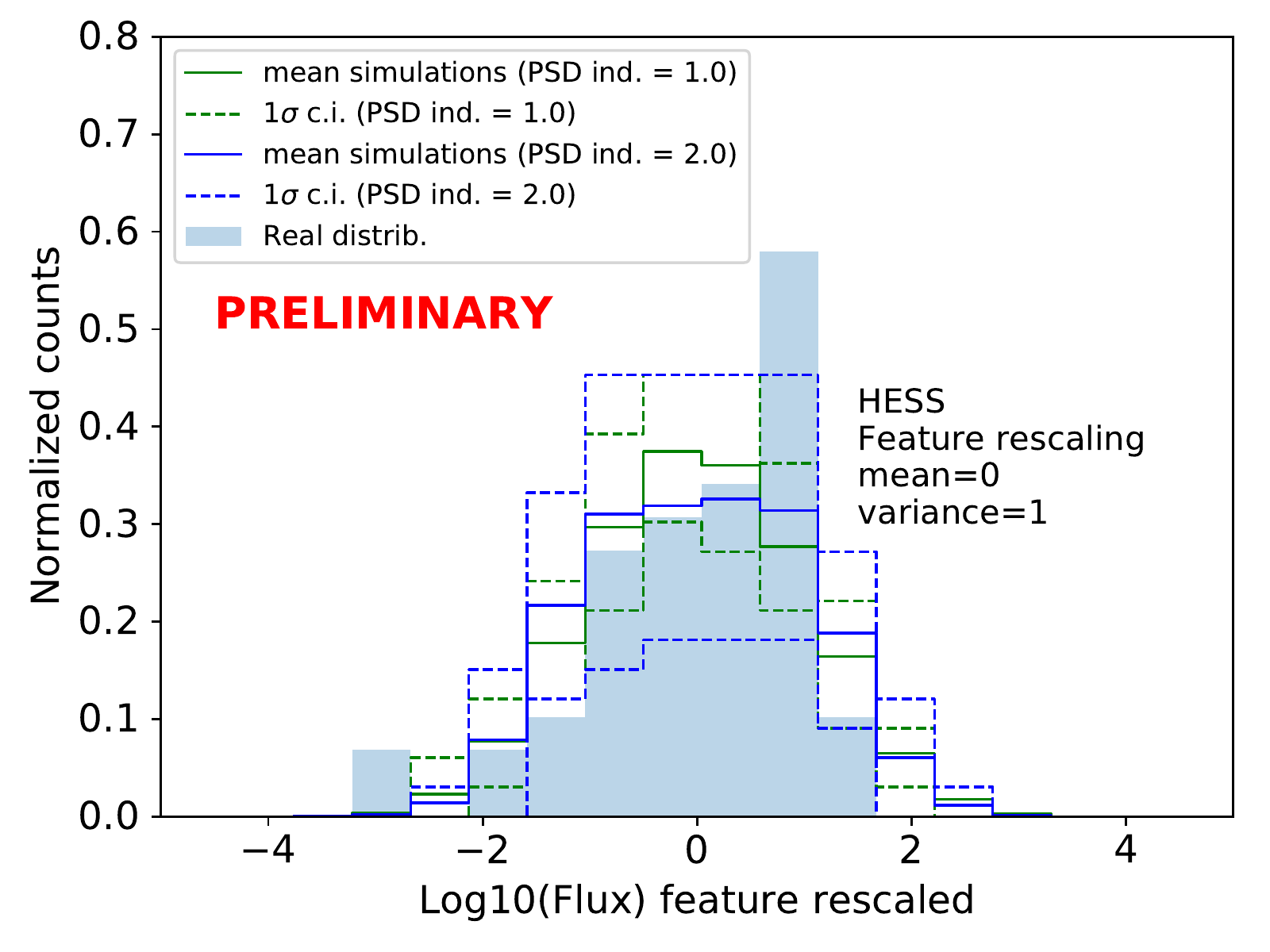}
\caption{Distribution of the rescaled flux for the H.E.S.S.\ 4-min-wise light curve compared with estimates from Gaussian ({\bf left panel}, 1000 simulations) and log-normal ({\bf right panel}, 1000 simulations) simulations. The colour codes are the same as in Figure~\ref{fig:PDFFermi}. In this case there seems to be only a slight preference for a log-normal behaviour with respect to a normal one, but neither distributions explain well the data}.
\label{fig:PDFHess}
\end{figure}

Finally, the H.E.S.S. data can probe the behaviour at the shortest variability timescales that can be examined with current data, thanks to the greater sensitivity of the instrument. The data used here come from the 3 nights of data centred in the peak of the flare. The histograms are shown in Figure~\ref{fig:PDFHess}. Also in this case, at the shortest timescales, we see hints of departures from Gaussian distributions of the flux. Even though also the log-normal one cannot explain properly the distribution of the data.

As for the other datasets, we report the \emph{p}-values for the various tests and hypothesis. Using~the SW test the \emph{p}-value for the normal case is $4\times10^{-4}$, while in for log-normal hypothesis $10^{-3}$. In~the comparison with the simulations we obtain \emph{p}-values $0.38_{-0.19}^{+0.27}$ and $0.63_{-0.19}^{+0.24}$ for normal and log-normal~respectively.

\section{Discussion and Conclusions}
\vspace{-6pt}

\subsection{Mrk~501 Results}
The data available for this study seem to confirm the trend seen in previous works: The variability of the very high energy gamma-ray emission shows strong departure from a Gaussian behaviour, pointing towards a preference of models that predict log-normal behaviour in the flux distribution especially when long term, unbiased datasets are available.

The fact that at GeV energies instead we cannot recognize clearly this trend leaves us with different possible explanations.
One of the possibilities is that the Mrk~501 could present a Gaussian behaviour in this energy range when considering long ($\sim$monthly) timescales, while on shorter timescales the trend would look more similar to what we see in the TeV regime. Unfortunately, the testing of this hypothesis is particularly challenging given the lack of sufficient statistics at GeV energies to realize light curves on timescales shorter than 1 month.
If instead the Gaussian behaviour is only present at low energies becoming log-normal in the TeV regime, this might be an indication of a different origin of the variability in the two energy ranges and hence a different origin of the emission. There is a degeneracy in the timescale-energy space that cannot be broken with the present data.

Recent preliminary studies done by the FACT Collaboration over a much longer data set on this source, reconstruct the flux distribution of Mrk~501  as a superposition of a Gaussian component, which~contributes mostly to explain the low state of the source, plus a log-normal one which is responsible for the bulk of the variability seen at TeV energies \cite{DornerFACT}. These long TeV light curves might be the key to disentangle timescales and energy effects.

Another possibility is offered by the model proposed in \cite{2018MNRAS.480L.116S}. The lack of log-normal distribution at GeV energies would be naturally accounted by the perturbations in the acceleration and/or escaping timescales. Furthermore, with tuning of the parameters in the scenario of perturbation on the escape time, it might even be possible to accommodate different shapes of flux distributions which are not exactly Gaussian or log-normal.

\subsection{The Importance of Unbiased Monitoring}
Having a strong handle on all the statistical properties of the variability phenomenon in blazars will be a powerful tool in the understanding the origin of the gamma ray emission in this sources.
The case of Mrk~501 shows how much we can benefit from the synergy of combined observations of different gamma-ray telescopes and the importance of monitoring programs. The more in depth we go in the study of well-known sources, the better we can understand and try to extrapolate properties to other objects \cite{2017NewAR..78...16T}. In this regard, the observation strategy adopted by the FACT telescope is of great relevance, because it provides data on a set of selected sources which are continuously observed, in an \emph{unbiased} way, regardless of their activity status. Hopefully, in the coming years, the FACT program will benefit from the presence of other similar telescopes, like M@TE, which will join this monitoring strategy and extend the continuous coverage further \cite{2017AIPC.1792g0007D}.
When considering monitoring activities at TeV energies, it is worth mentioning the non-imaging gamma-ray detectors like the HAWC (High Altitude Water Cherenkov) Observatory. Detecting directly the shower particles reaching the ground, this type of experiment has the advantage of having a duty cycle of almost 100\%, not being bound by the presence of the Sun or the Moon. The disadvantage here is that the energy threshold is typically higher compared to IACTs. The low sensitivity at sub-TeV energies makes the observation of extragalactic objects more difficult, unless they are nearby and bright. Currently, the only two extragalactic sources seen by HAWC are the blazars Mrk~421 and Mrk~501 \cite{2017ApJ...843...40A}. Despite this, HAWC is already producing interesting results on the study of these two AGNs \cite{2017ApJ...841..100A}, being able to perform a daily monitoring.
For the high sensitivity observations on short timescales, for the moment we rely on the classic IACT arrays (MAGIC, H.E.S.S., VERITAS), but in a future not too far CTA will start its operation with a planned sensitivity one order of magnitude higher, meaning that it will be potentially possible to probe sub-minute timescales and have availability of high precision data on a large energy range \cite{2017arXiv170907997C} for the study of variability phenomena. For continuous monitoring, larger~HAWC-like instruments are in construction \cite{LHAASO} or planned\footnote{\url{https://www.sgso-alliance.org}} \cite{SGSO}, which would provide an excellent synergy with CTA.
The future of the GeV field is a bit more uncertain with not yet a designed successor of the {\it Fermi}-LAT telescope in this energy band. Proposed missions for the MeV energy range like {\it eASTROGAM}\footnote{\url{http://eastrogam.iaps.inaf.it/}} and {\it AMEGO}\footnote{\url{https://asd.gsfc.nasa.gov/amego/}} would still have good sensitivity in the GeV energy range, but they are still in the proposal stage and the timescale could be as late as mid-2030s.
Hence, it would be important to keep the {\it Fermi} satellite running for as long as possible given its fundamental importance for variability studies in the GeV regime.

%
%
\vspace{6pt}

\authorcontributions{C.R. performed the {\it Fermi}-LAT analysis and wrote the majority of this manuscript; N.C.~performed the H.E.S.S. analysis, produced the simulated light curves and contributed to this paper; D.D. and M.B. performed the FACT analysis and contributed to this paper; A.T. reviewed the analysis methodology and contributed to this paper.}

\acknowledgments{
{\bf FACT Collaboration}: The important contributions from ETH Zurich grants ETH-10.08-2 and ETH-27.12-1 as well as the funding by the Swiss SNF and the German BMBF (Verbundforschung Astro- und Astroteilchenphysik) and HAP (Helmoltz Alliance for Astroparticle Physics) are gratefully acknowledged. Part of this work is supported by Deutsche Forschungsgemeinschaft (DFG) within the Collaborative Research Center SFB 876 ``Providing Information by Resource-Constrained Analysis'', project C3. We are thankful for the very valuable contributions from E. Lorenz, D. Renker and G. Viertel during the early phase of the project. We~thank the Instituto de Astrofísica de Canarias for allowing us to operate the telescope at the Observatorio del Roque de los Muchachos in La Palma, the Max-Planck-Institut für Physik for providing us with the mount of the former HEGRA CT3 telescope, and the MAGIC collaboration for their support. {\bf HESS Collaboration:} The support of the Namibian authorities and of the University of Namibia in facilitating the construction and operation of H.E.S.S. is gratefully acknowledged, as is the support by the German Ministry for Education and Research (BMBF), the Max Planck Society, the
German Research Foundation (DFG), the Helmholtz Association, the Alexander von Humboldt Foundation,
the French Ministry of Higher Education, Research and Innovation, the Centre National de la
Recherche Scientifique (CNRS/IN2P3 and CNRS/INSU), the Commissariat à l’énergie atomique
et aux énergies alternatives (CEA), the~U.K. Science and Technology Facilities Council (STFC),
the~Knut and Alice Wallenberg Foundation, the National Science Centre, Poland grant no. 2016/22/M/ST9/00382, the~South African Department of Science and Technology and National Research Foundation, the University of Namibia, the~National Commission on Research, Science \& Technology of Namibia (NCRST), the~Austrian Federal Ministry of Education, Science and Research and the Austrian Science Fund (FWF), the~Australian Research Council (ARC), the~Japan Society for the Promotion of Science and by the University of Amsterdam. We appreciate the excellent work of the technical support staff in Berlin, Zeuthen, Heidelberg, Palaiseau, Paris, Saclay, Tübingen and in Namibia in the construction and operation of the equipment. This work benefited from services provided by the H.E.S.S. Virtual Organisation, supported by the national resource providers of the EGI Federation.
}


\conflictofinterests{The authors declare no conflict of interest.}






\reftitle{References and Notes}




\end{document}